# Resolving Length Scale Dependent Transient Disorder Through an Ultrafast Phase Transition

Jack Griffiths[1,*], Ana Flávia Suzana[1], Longlong Wu[1], Samuel D. Marks[2], Vincent Esposito[3], Sébastien Boutet[3], Paul G. Evans[2], J. F. Mitchell[4], Mark P. M. Dean[1], David A. Keen[5], Ian Robinson[1], Simon J. L. Billinge[1,6], Emil S. Bozin[1,*]

[1] Condensed Matter Physics and Materials Science Division, Brookhaven National Laboratory, Upton, NY 11973, USA
[2] Department of Materials Science and Engineering, University of Wisconsin, Madison, WI 53706, USA
[3] SLAC National Accelerator Laboratory, 2575 Sand Hill Road, Menlo Park, California 94025, USA
[4] Materials Science Division, Argonne National Laboratory, Lemont, Illinois 60439, United States
[5] ISIS Neutron and Muon Source, STFC Rutherford Appleton Laboratory, Harwell Campus, Didcot, Oxfordshire OX11 0QX, United Kingdom
[6] Department of Applied Physics and Applied Mathematics, Columbia University, New York, New York 10027, USA
*Corresponding Authors

**Material functionality can be strongly determined by structure extending only over nanoscale distances. The pair distribution function presents an opportunity to shift structural studies beyond idealized crystal models and investigate structure over varying length scales. Applying this method with ultrafast time resolution has the potential to similarly disrupt the study of structural dynamics and phase transitions. Here, we demonstrate such a measurement of $CuIr_2S_4$ optically pumped from its low temperature Ir-dimerized phase. Dimers are optically suppressed without spatial correlation, generating a structure whose level of disorder depends strongly on length scale. The re-development of structural ordering over tens of picoseconds is directly tracked over both space and time as a transient state is approached. This measurement demonstrates both the crucial role of local structure and disorder in non-equilibrium processes and the feasibility of accessing this information with state-of-the-art XFEL facilities.**

The development of materials with specialized and highly efficient properties increasingly relies on complex local structures that stray from the ideal of a perfect crystal[1–3]. In particular, advancements in electronics technology drive a need for materials that switch between distinct states: either electrical (e.g., memristors[4], ferroelectrics[5]), magnetic (e.g., ferromagnets[6], anti-ferromagnets[7]) or structural (e.g., charge density wave states[8]). A key example is the metal-insulator transition[9,10]. It is well established that local structure plays a central role in many equilibrium phase transitions driven by the competition of energy and entropy[11–13]. Some non-equilibrium phase transitions can be triggered on demand using ultrafast laser pulses. Local structure has also been implicated in these transitions[14,15], but this is less understood due to a lack of appropriate means to quantify length scale dependent local disorder in these ultra-fast transient states.

In studies at equilibrium, the pair distribution function (PDF) plays an integral role in characterizing locally broken structural symmetry and structural disorder[16,17]. This function of interatomic distances in the scattering material, generated through a Fourier transform of an appropriately normalized scattering pattern, is a quantitative and easily interpretable probe for atomic structure on all length scales from local to bulk. X-ray free electron laser (XFEL) facilities offer a key opportunity to apply the PDF technique to picosecond structural dynamics, such as phase transitions, using high brilliance 100 fs X-ray pulses. This would represent a x10$^9$ increase in temporal resolving power compared to a typical synchrotron experiment that does not severely compromise brilliance with slicing or fast shutter methods[18]. In comparison to electrons, which can also be generated in ultra-short pulses, X-rays scatter kinematically and can generate quantitatively reliable PDFs.

Here, we demonstrate the feasibility of XFEL ultrafast-PDF (uf-PDF) to track an optically pumped phase transition in $CuIr_2S_4$ (CIS). This measurement shows that while the local atomic structure transitions in less than a picosecond, the average structure on length scales longer than a unit cell continues to strongly evolve for tens of picoseconds as long-range order between local regions is established. Although the pumping process produces a transition between two ordered phases, this measurement tracks the pivotal role of disorder through the transition itself.

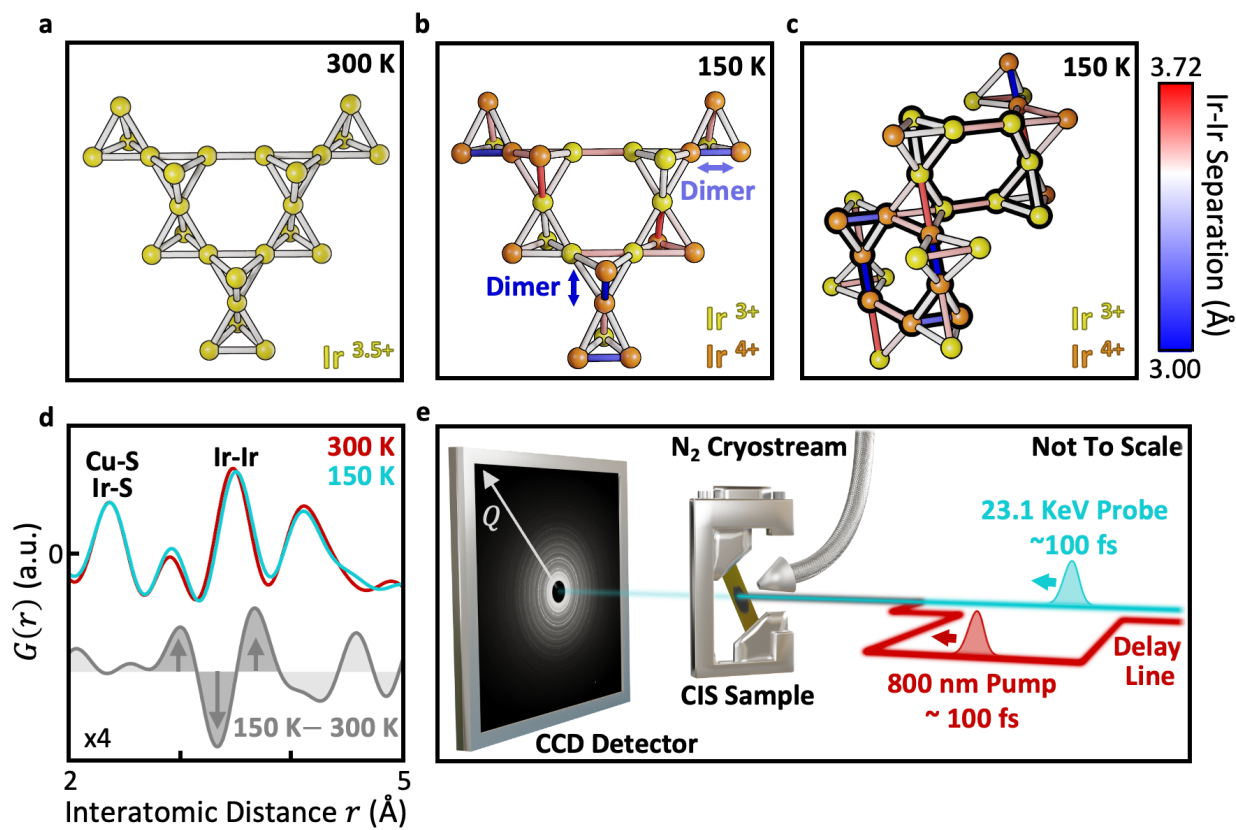

**Figure 1 | $CuIr_2S_4$ Ir Dimerization.** a) Portion of the undistorted pyrochlore Ir substructure in the 300 K CIS structure described in space group symmetry $Fd\bar{3}m$. Structure refined from experimental data. b) Portion of the CIS Ir substructure at 150 K as described in triclinic space group $P\bar{1}$. Charge/orbital ordering and spin dimerization separates Ir-Ir bonds into significantly shortened (blue) and lengthened (red), or largely unaltered (grey) groups. Dimerized bonds run along two distinct [110]-type cubic directions (examples marked by arrows). Structure refined from experimental data. c) Different portion and view of the same 150 K CIS Ir substructure, emphasizing topologically equivalent bi-capped hexamers containing either $Ir^{3+}$ or $Ir^{4+}$ ions (bold outlines). d) Ir dimerization creates an 'M'-shaped differential PDF signature at the length scale of an Ir-Ir bond

(arrows). PDFs measured at the LCLS XFEL in equilibrium. Differential magnified x4 for clarity. e) Experimental schematic of XFEL pump-probe uf-PDF measurements of low temperature CIS. Pump delay line is able to adjust the relative pump-probe arrive time at the sample from -20 ps (maximum pump delay) to 100 ps (minimum pump delay).

CIS is a spinel material exhibiting a metal-to-insulator phase transition (generally described as Peierls-like) upon cooling through 226 K[19]. Above this temperature, it consists of a cubic $Fd\bar{3}m$ unit cell with Ir atoms, the dominant X-ray scatterers, forming a pyrochlore substructure of regular tetrahedra with edge length ~3.5 Å (Fig. 1a). This generates a strong PDF peak at this interatomic distance. Below 226 K, the Ir undergo charge and orbital ordering as their effective 3.5+ charge disproportionates to a nominally 3+/4+ state. Simultaneously, spin dimerization (Fig. 1b,c) shortens (dimers) and lengthens some Ir-Ir distances with a separation in length of ~0.7 Å. The resulting 'M'-shaped signature in the difference of PDF profiles across the transition is significant enough to be well resolved with the limited resolution of an XFEL measurement (Fig. 1d). The dramatic loss in symmetry through this transition is reflected by a new triclinic $P\bar{1}$ unit cell[19]. The Ir dimerization can be described by chains running along two distinct [110]-type cubic directions[20] (Fig. 1b, arrows) or, alternatively, by two topologically identical 8 atom bicapped hexamers (referred to from here as Ir octamers)[19]. These consist either of $Ir^{4+}$ (containing dimerized bonds) or $Ir^{3+}$ (containing non-dimerized bonds) ions and together tile 3D space (Fig. 1c, bold outlines). In 2019, high-resolution synchrotron PDF was used to investigate how the local structure of CIS harbingers the bulk phase transition[21]. While the metallic phase is cubic on average, each Ir-Ir bond dynamically fluctuates by <0.1 Å due to an orbital degeneracy lifting (ODL) precursor effect that reduces the local symmetry to tetragonal $I4_1/amd$ (a subgroup of $Fd\bar{3}m$). These ODL dimers, which are correlated over increasing distances as the phase transition is approached, cast doubt on the Peierls mechanism of the metal-to-insulator transition and suggest a greater complexity that is unlikely to be understood from equilibrium structural measurements.

Ultrafast reflectivity studies[26,27] have shown that dimerized CIS responds to optical pumping with reflectivity decreasing and recovering over sub-picosecond and tens of picoseconds timescales respectively. Using multi-pulse techniques, it has been argued that the (so far unidentified) pumped phase represents a new transient structure and not a return to the high temperature $Fd\bar{3}m$ state[26]. The pumped phase is weakly conducting, suggesting removal of strong dimerization, and could therefore be speculatively related to an ordered variant of the ODL state. As the removal of strong dimerization in CIS would generate a large enough PDF signal to be clearly resolvable in an XFEL measurement (Fig. 1d), this pumped transition is ideal for appraising the uf-PDF technique while simultaneously examining how CIS transitions between insulating and conducting states. Note that dimerized CIS is also known to be sensitive to continuous irradiation by UV or X-ray photons over hundreds of milliseconds[22–25]. In this markedly distinct regime of photon energies and peak fluences, the X-ray Bragg scattering signature of dimers is removed. However, PDF has shown that only long-range dimer order is destroyed while the dimers themselves persist locally[23].

To investigate this optically driven transition, a layer of powdered CIS was pumped at 150 K using an 800 nm laser pulse. The pumped sample was stroboscopically probed using X-ray pulses using a transmission scattering geometry at the MFX beamline of the LCLS XFEL facility (Fig. 1e, see Methods). A scattering momentum transfer range from 1.6 Å$^{-1}$ to 12.6 Å$^{-1}$ was achieved using 23.1 keV photons. This can be compared to other current XFEL scattering and PDF (without time resolution) measurements achieving only maximum momentum transfers of 5 – 8 Å$^{-1}$ that severely limit real space resolution and degrade the information needed for quantitative analysis[14,28,29]. A reference (unpumped) measurement of the sample was also taken at the same temperature at the Advanced Photon Source synchrotron with a larger maximum momentum transfer of 23 Å$^{-1}$.

**Results**

Initial observations regarding the structural response to the pump laser can be made in reciprocal space. The reduced structure factors $F(Q)$, where $Q$ is the scattering momentum transfer, and the difference curves $\Delta F(Q)$, where an averaged unpumped reference is subtracted to amplify more subtle changes, are shown as a function of pump-probe delay in Figure 2a,b. There is a clear abrupt change in the patten upon crossing 0 ps pump-probe delay, indicating a structural response. These changes then evolve smoothly in time over the measured $Q$ range and do not begin to reverse in the 100 ps measurement time, indicating a lifetime at least this long for the pumped structural phase. These observations suggest that the pumping process can be described in terms of three structures: the *Dimerized* structure before pumping (which is well characterized), the *Prompt* structure immediately upon pumping (characterized in this work) and the *Transient* structure that the material relaxes to over tens of picoseconds (also characterized here). Importantly, no significant response – including heating induced lattice expansion – was observed when pumping the high temperature phase (Supplementary Figure 1, 2).

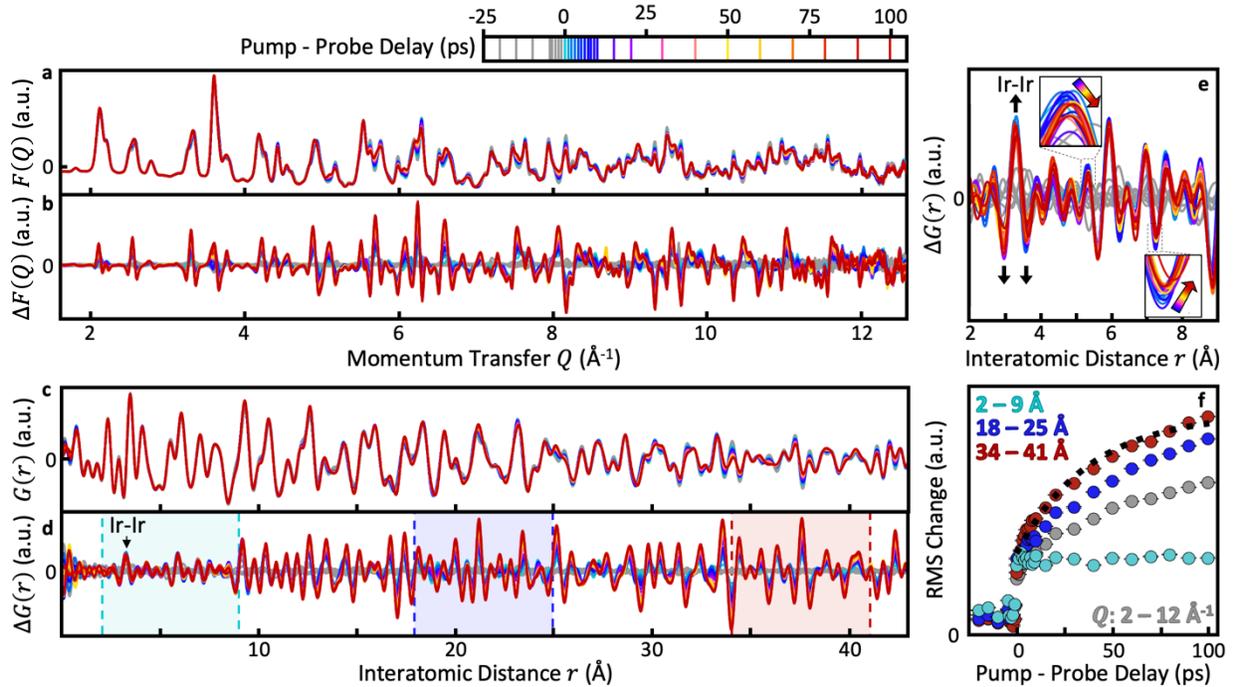

**Figure 2 | Structural Response to Laser Pump.** a) Reduced structure factor $F(Q)$, a linear function of the diffraction pattern that magnifies high $Q$ features, with varying pump-probe delay. All negative delay signals (unpumped measurements) are colored grey. v) $\Delta F(Q)$ subtracting the average of {-20,-15,-10} ps to emphasize differences due to laser pump. Y-axis magnified x3.4 relative to $F(Q)$. c) PDF $G(r)$ with varying pump-probe delay. d) $\Delta G(r)$ subtracting the average of {-20,-15,-10} ps to emphasize differences due to laser pump. Y-axis magnified x3.4 relative to $G(r)$. e) $\Delta G(r)$ over sub-nanometer length scales showing a sub-ps abrupt change. The signature of Ir-Ir dimer suppression, the inverse of the 'M' in Figure 1d, is marked with black arrows. Insets: Smaller changes to the PDF continue with increasing delay. f) Root-mean square (RMS) of $\Delta G(r)$ calculated over equally sized ranges of interatomic distance, showing the same initial prompt response at 0 ps but a difference in further structural evolution at local and non-local length scales. $\Delta G(r)$ RMS over 34 – 41 Å range follows exponential behavior with (38 ± 6) ps characteristic time (black dashed line). RMS of $\Delta F(Q)$ is also shown (grey) representing all accessible length scales. $\Delta F(Q)$ normalised to match negative delay values for $\Delta G(r)$ (representing the noise level). Error bars indicate uncertainty due to photon shot noise only.

PDFs $G(r)$, where $r$ is interatomic distance, are generated from sine Fourier transforms of the reduced structure factor (Fig. 2c,d), where the finite range of measured $Q$ applies a well understood convolution that broadens peaks and introduces termination ripples. Despite these termination artefacts, analyzing the data in real space has several key advantages. First, subtle and/or broadband changes to diffuse scattering are converted to changes in the positions and shapes of peaks that are easier to identify, interpret and model. Second, focusing on different regions of a PDF reveals how the average atomic structure varies over different length scales. This provides information on structural ordering. Here, the difference curves $\Delta G(r)$ display the 'W' signature that some strong Ir-Ir dimers are removed by the laser pump (Fig. 2e [arrows] opposite to the 'M' feature in Fig. 1d [arrows]). The central positive peak of this 'W' along with the two adjacent negative valleys represents a reduction in the spread of Ir-Ir distances as they shift towards a central value. As this signature is fully formed at 1 ps pump-probe delay, this strong dimer suppression must occur on femtosecond timescales and is not probed temporally here. This is consistent with optical dimer removal in $VO_2$, for example, which occurs over 100 fs[14]. The shape of the sub-nanometer response bears a strong qualitative agreement to a simple calculation of multiple unit cells in which either a single or all Ir-Ir dimers are lengthened by 0.4 Å (see Supplementary Note 1 for details).

The >1 ps time-evolution of the pumped PDFs strongly depends on length scale $r$. Following the laser pump, the features in the difference curve $\Delta G(r)$ are initially similar in scale at all $r$ (Fig. 2d). With continuing delay, these scales evolve differently above and below ~9 Å – the length scale of one unit cell / octamer (Supplementary Figure 3). While the PDF below 9 Å, and therefore the distribution of nearest neighbor atomic distances, could be tentatively argued to continue to subtly evolve with delay just above the noise level (Fig. 2e insets), any changes are dwarfed by the initial abrupt response. This is clear from the constant Root Mean Square (RMS) of $\Delta G(r)$ from 2 – 9 Å, reducing the information in $\Delta G(r)$ to a simple magnitude of the PDF change (Fig. 2f). In contrast, this same metric at longer length scales (18 – 25 Å and 34 – 41 Å) features a similar initial step change followed by strong growth incomplete within 100 ps. This implies that the pumped *Prompt* and *Transient* phases are the same on local length scales but distinct over length scales averaging over multiple unit cells. Both differ from the starting *Dimerized* phase at all length scales. The longer timescale increase in the RMS metric over 34 – 41 Å follows an exponential time constant of $(38 \pm 6)$ ps.

A picture emerges from these real-space observations. The structural changes instigated by the optical removal of Ir-dimers (a stochastic process) are initially uncorrelated between local regions of the sample. That is, in the *Prompt* phase, the spatial arrangement of bond lengths differs from one local region to the next. Internal strain imposed by this disorder likely drives the re-development of a non-equilibrium long-range order over time. This alters the PDF over longer length scales as the *Transient* phase is approached while leaving the local PDF over sub-unit cell length scales largely unchanged. A minor evolution of the local PDF with delay could be permitted as external strain on each unit cell reduces with decreasing disorder. Note that the observation of the removal of strong dimers precludes the pumped state from being the same as that reached under continuous UV or X-ray irradiation in which local dimerization is preserved[23].

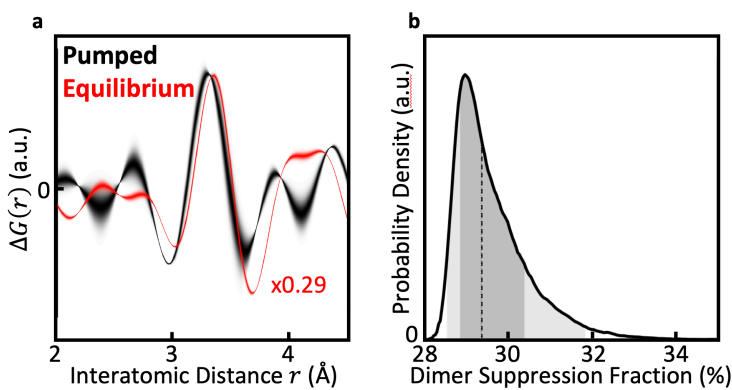

**Figure 3 | Dimer Suppression Signature.** a) Experimental $\Delta G(r)$ for both the pumped (black) and thermally driven equilibrium (red) transitions, showing the dimerized PDF subtracted from the un-dimerized PDF, over the low $r$ range containing the W-signature centered at 3.5 Å of strong dimer removal. The pumped data averages over all positive pump-probe delay times with line thickness indicating uncertainty (Supplementary Note 2). The equilibrium data is scaled by x0.29 so that the signatures approximately match in scale. b) Uncertainty on PDF normalization is propagated to a Probability Density Function representing the scaling between the pumped and equilibrium dimer suppression signatures (Supplementary Note 2). Dashed line indicates the median probable value of 29.4%. Shaded areas indicate the central 68% and 95% probability intervals.

The average W signature centered at 3.5 Å in $\Delta G(r)$ can be compared to the same signature generated by the equilibrium (thermally driven) transition in which all strong dimers are removed. While qualitatively very similar, the thermally driven signature must be scaled by ~0.29x to match the pumped signature intensity. This indicates that ~29% of probed dimers in the pumped sample are suppressed. As the pump laser is expected to penetrate 40 nm into each ~0.7 μm powder grain[26], we would expect only a maximum of ~4% of probed dimers to be suppressed depending on the fraction of dimer suppression within the pumped volume. This suggests that either a) the characterization of the powder was inaccurate and the average grain size is smaller than determined by confocal microscopy (Supplementary Figure 4) or b) the transition occurs within a greater depth with energy carried by, for example, non-thermal photoexcited electrons[30,31]. Although the pumped phase fraction cannot be unambiguously determined, there is a lower bound of ~29% on the proportion of dimers suppressed within the pumped volume.

The pumped signature can also be compared to a simple 'small box' model, a numerically generated PDF of a perfect crystal (with experimental effects such as termination artefacts reproduced) that is typically compared to an experimental PDF over a certain interatomic distance range. These models parameterize, with symmetry constraints, the atomic positions within a unit cell and the Atomic Displacement Parameters (ADPs) that smear these positions both due to thermal motion (temporal variance) and any disorder between local regions (spatial variance). For describing this narrow $r$-range, we assume a simple model where the Ir substructure linearly transitions between the dimerized and un-dimerized configurations (Supplementary Note 3). Using this model, we find that the resolution and signal-to-noise ratio is not high enough to distinguish between Ir dimers being significantly weakened or entirely removed (Supplementary Note 3). This model, implicitly assuming that all strong dimers in the pumped volume are suppressed by the pump, suggests a pumped phase fraction of ~20%. The small shift between the W signatures of the pumped and equilibrium transition (Figure 3a) could indicate an inflated Ir ADP due to the pumping (Supplementary Note 3) or simply represent a small (~3%) inaccuracy in the intensity normalization of the PDFs.

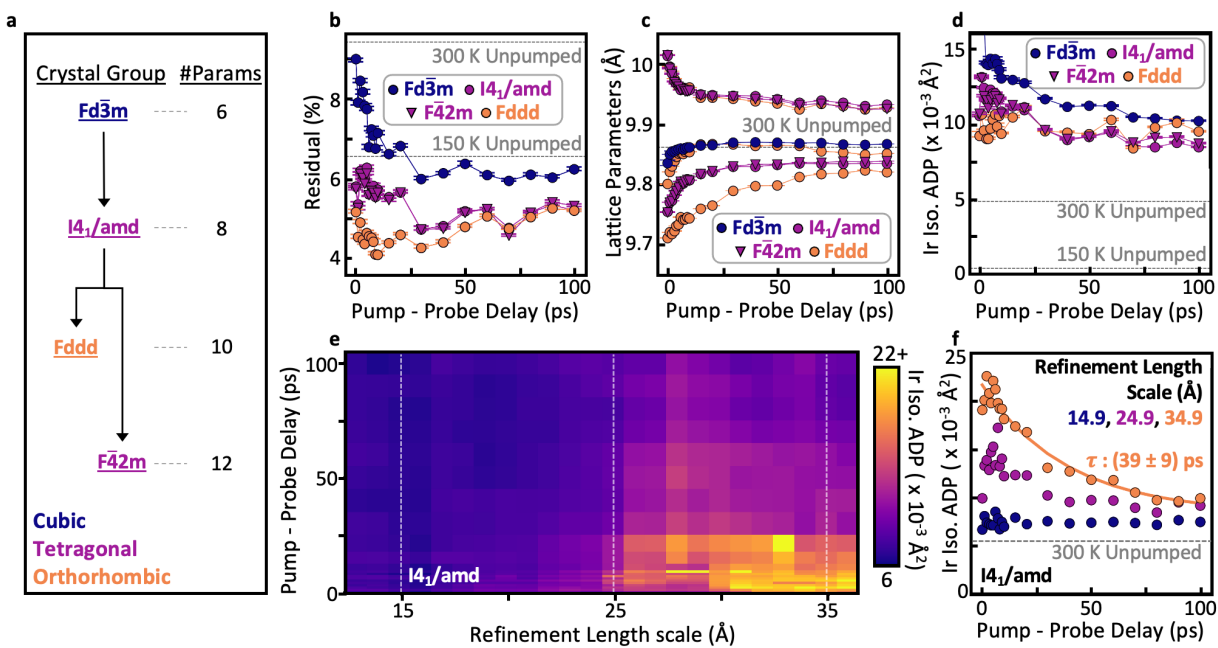

**Figure 4 | Models of Non-Local Structure** a) Crystal space groups used to model the average (8.8 – 40 Å) length scale of the pumped PDFs. Arrows denote a group-subgroup relation. Each subgroup exposes more refinable structural parameters (denoted #Params). Each model fit also includes four parameters capturing properties of the experimental set-up. b) Residual $rw$ versus pump-probe delay. Dashed lines indicate reference $rw$ for unpumped models at 300 K (Fd$\bar{3}$m) and 150 K (P$\bar{1}$). c) Lattice parameters versus delay. Dashed line indicates reference lattice parameter for 300 K equilibrium measurement. d) Ir isotropic ADPs versus delay. Dashed lines indicate reference ADPs for unpumped 300 K and 150 K measurements. All error bars indicate uncertainty due to photon shot noise only. See Supplementary Fig. 11, 12 for example fit. e) I4$_1$/amd isotropic Ir ADPs refined over a sliding window of width 8.2 Å to generate a spatial-temporal disorder map. Refinement Length Scale indicates the center of the refinement windows. f) ADPs with delay for windows at three representative length scales. Longer length scale ADPs decrease over time with a time constant of ~ 40 ps while shorter length scale ADPs remain constant. Delay times are marked by dashed lines in e).

The small box methodology is also applied to non-local length scales (8.8 – 40 Å). In this case, the larger $r$ range allows meaningful 3D structural information to be extracted. The experimental PDFs are modelled as a sum of a calculated *pumped* phase and an assumed static *unpumped* phase that is here defined as proportional to the measurement at -20 ps pump-probe delay. Including experimental data as a model component will artificially reduce fit residuals $rw$ (see Methods for definition) as this component will, by definition, capture any systematic errors/artefacts present in the data. Accordingly, more focus must be applied to relative differences between pump-probe delays and between different structural models. In the high temperature equilibrium phase, weak fluctuating dimers exist locally but are uncorrelated over multiple unit cells to average out to a cubic structure in space group $Fd\bar{3}m$ at non-local length scales. Applying the $Fd\bar{3}m$ unit cell here leads to a larger $rw$ over the first 25 ps delay, indicating underfitting of short delay data and a lack of model complexity (Fig. 4b). Therefore, the averaged pumped structure is not $Fd\bar{3}m$ but a new phase in agreement with a previous reflectivity study[26]. These higher residuals are reduced by using a tetragonal structure in space group $I4_1/amd$ (a sub-group of $Fd\bar{3}m$). This is the unit cell used to describe the high temperature local ODL structure. At low delays, residuals can be further improved by reducing the symmetry again to an orthorhombic structure in Fddd (a sub-group of $I4_1/amd$). The orthorhombic and tetragonal residuals converge at higher delays. Note that both $I4_1/amd$ and Fddd descriptions are simple, high symmetry models with Ir atom positions fixed within the unit cell and any change in Ir-Ir distances are dependent on the lattice parameters. Further increasing the complexity of the model, for example by using a tetragonal $F\bar{4}2m$ unit cell that allows some limited changes to Ir atom positions independent of the unit cell parameters, offers no further improvement in $rw$ and begins to enter the regime of overfitting. While higher resolution data, taken with a future higher energy XFEL or novel detector geometry, may be able to successfully resolve a lower symmetry model in the future, this data is therefore best described non-locally by an orthorhombic unit cell at lower delay times (the *Prompt* phase) and a tetragonal unit cell at longer delay times (the *Transient* phase). The apparent increase in symmetry with delay at non-local length scales reflects an evolution in the longer-range order between the local, strong dimer removed, regions and is consistent with the model-independent analysis. Note that these models provide pumped phase fractions of 20-25 % (Supplementary Figure 13), consistent with the above analysis of the local PDF.

Modelling provides access to information that is not obvious from the raw data alone. For example, both $I4_1/amd$ and Fddd unit cells involve a new well defined exponential timescale of $(12 \pm 1)$ ps governing the lattice parameters (Fig. 4c, interpreted below). With the knowledge to carefully search for it, this timescale can be identified in the raw data where it tracks the subtle shifting of some peak positions in $\Delta G(r)$ (Supplementary Figure 14). Modelling also provides access to ADPs. For the $I4_1/amd$ unit cell, the Ir ADPs decrease over tens of ps (Fig. 4d) indicating a reduction in spatial disorder and/or thermal motion as expected. By re-refining this model over a narrower sliding window of $r$, ADPs can capture how disorder varies as a function of length scale. The model is re-refined using a sliding window of width 8.2 Å – a window size that did not risk overfitting. While unit cell structure does not significantly depend on the refinement range (Supplementary Figure 15), the resulting Ir ADPs show a strong dependence (Fig. 4e). This dependence does not correlate with other model parameters and is therefore not a result of overfitting (Supplementary Figure 15). As a similar strong dependence is also observed for the more complex tetragonal $F\bar{4}2m$ model, this is also not a consequence of model underfitting. Below ~25 Å (i.e., 2–3 unit cells), the Ir ADPs are easily attributed to thermal effects (~0.006 Å$^2$) and show no significant delay dependence indicating insignificant laser heating. At larger length scales, the ADPs initially increase by over 350% (to ~0.022 Å$^2$) before decreasing with delay with a characteristic time of $(39 \pm 9)$ ps (Fig. 4f). This tracks the ~40 ps time scale of the PDF RMS metric (Fig. 2f). At short delays (*Prompt* phase), the pumping of regions separated above a critical distance (seemingly 2-3 unit cells) can be considered independent and their spatial arrangement of bond lengths uncorrelated. This inflates ADPs forced to capture this disorder. The resulting strain between local regions drives lattice relaxation and a (39 $\pm$ 9) ps fall in the disorder-related ADP component as the *Transient* phase is approached.

**Discussion**

The evolution of non-local (8.8 – 40 Å) pumped CIS from the *Prompt* to the *Transient* structures contains distinct ~12 ps and ~40 ps timescales governing lattice parameters and Ir ADPs respectively. The existence of two seemingly independent timescales can be tentatively explained. The average unit cell averages over the local structure at every point in the pumped material. Upon initial pumping, there will likely be different local arrangements of atoms that give near-equivalent distributions of nearest-neighbor bond lengths. Our data is not sensitive to these different configurations. These different local structures average to the non-local unit cell, with the proportions of each configuration changing over the ~12 ps timescale. The proportions of local structural configurations gives little information on how spatially ordered they are with respect to one another, which is instead tracked by the ~ 40 ps timescale.

In summary, this uf-PDF investigation has shown that optical pumping suppresses strong Ir-Ir dimers in low temperature CIS. Despite transitioning between phases with long- and intermediate- range order respectively, this optically driven transition is fundamentally characterized by a period of high structural disorder. Above a critical distance (approximately 2 unit cells), the sub-picosecond suppression of Ir-dimers is uncoordinated with local regions taking on different spatial arrangements of bond lengths. We propose that internal strain drives a recovery of order over ~40 ps and an evolution in the average crystallographic structure. This timescale matches the recovery of reflectivity in previous studies[26,27]. The key power of PDF to explicitly isolate ranges of length scales for structural modelling has here allowed disorder, measured through ADPs, to be mapped over both length- and time-scales. This disorder mapping could be widely applied to other non-equilibrium systems.

In agreement with reflectivity studies[26], we find that pumped CIS is not driven back to the equilibrium room temperature phase. Instead, the non-local structure at longer delay times is best described using the I4$_1$/amd crystal space group that also describes the local ODL structure at room temperature. It is possible that this new phase is related to the ODL state with the fluctuating weak dimers now ordered over long length scales. In similar spinel structures, such as LiRh$_2$O$_4$[32], the metal-to-insulator transition involves first the ordering of fluctuating dimers and then spin dimerization at two distinct temperatures. In CIS, these processes are simultaneous. Speculatively, the non-equilibrium pumped phase may be this otherwise inaccessible (hidden) intermediate orbitally ordered state lacking spin singlet dimerization. Notably, this phase is distinct from the disordered dimer state achieved by UV, electron, and X-ray irradiation which preserves the dimers locally[22–25]. We caveat these conclusions by noting that the data, although currently the best available, is limited by a higher minimum $Q$, lower maximum $Q$ and broader $Q$-resolution than desirable. In a future experiment, it might be possible to distinguish between the symmetries of structural models more robustly and describe the pumped structure with a lower symmetry unit cell. These current resolution limitations prevent an in-depth structural analysis of the local (sub-unit cell) structure beyond the key observation that strong Ir dimers are optically suppressed.

Methodologically, this work has successfully demonstrated uf-PDF as a practical and powerful technique given a favorable sample with structural changes significant enough to be resolvable at current XFEL facilities. Despite the reduced PDF resolution and shot-to-shot consistency of an XFEL measurement compared to a synchrotron, these are not significant enough to obscure critical length-scale dependent structural information. Proven feasible, uf-PDF could be applied to better understand systems, such as $VO_2$[14,33] and $1T-TaS_2$[34], that also display transient disorder. This technique will only improve as experimental and data handling protocols are optimized and higher XFEL energies, possibly in the pipeline for the next decade, increase spatial resolution. As the picosecond time resolution of this experiment was set by precision expectations for the laser pump delay line, there is no reason why this now-demonstrated technique couldn't be pushed far into the femtosecond regime. As a complement to studies of diffuse scattering that remain in reciprocal space[14], uf-PDF is highly quantitative through straightforward comparison to structural models.

## Methods

### Sample Synthesis

*$CuIr_2S_4$ was prepared by solid state reaction in evacuated quartz ampoules. Stoichiometric quantities of the metals and elemental sulfur were thoroughly mixed, pelletized, and sealed under vacuum. The ampoules were slowly heated to 650–750 °C and soaked for several weeks with intermediate regrinding and pelletizing. The reaction was deemed complete when no further changes in x-ray powder diffraction scans were observed. The product was found to be a single spinel phase.*

### Measurements

*A 2-3 µm layer of powdered CIS (grain size ~ 1 µm), spread onto Kapton tape, was measured using a transmission pump-probe geometry at the MFX beamline of the LCLS XFEL source. Full Debye-Scherrer rings of scattered X-rays were collected using a Rayonix MX340 CCD positioned ~70 mm downstream of the sample (i.e. a Rapid Acquisition PDF setup[35]) at an acquisition rate of 30 shots/second. An 800 nm ~100 fs laser pulse from a Coherent Vittara was used to pump the sample with 41 µJ of energy over a 400 µm Full Width at Half Maximum Gaussian spot. A 23.1 KeV X-ray beam probed the sample in ~100 fs pulses. The probed area, a 300 µm top-hat spot, was smaller than the pumped region to minimize any effects from the spatial distribution of pump energy. The sample was cooled using a $N_2$ cryostream at a nozzle temperature of 150 K. Each delay time was measured stroboscopically and averaged over hundreds of pump-probe cycles. The pump-probe delays were measured in a pseudo-random order to ensure no accruing permanent structural changes (i.e. damage) from repeated pumping. Background measurements were acquired of both bare Kapton tape and air scatter, and dark measurements were taken without application of the X-ray probe pulse.*

### PDF Generation

*The 2D diffraction image associated with each pump-probe delay time was the average of hundreds of individual stroboscopic measurements. The number of averaged images was not constant with delay due to the varying number*

of 'bad shots' filtered out (where the measurement fails due to a mechanical or electrical fault). Background measurements of air scatter and bare Kapton tape were removed. The images were normalized for the polarization dependance of the detector's detection efficiency and for any variation in detection efficiency across the detector (the 'flat field'). An unexpected intensity was noted that was broadband in $Q$ and did not vary in intensity around the Debye-Scherrer rings (as expected from the polarization dependance). This was likely the result of multiple-scattering events due to the sample thickness. As this did not vary in intensity around the rings like the single-scatter data, it could be removed post-hoc. The 2D images were converted to 1D diffraction patterns using the pyFAI software[36].

Reference unpumped measurements from the APS synchrotron with higher maximum momentum transfer magnitude $Q \sim 23$ Å$^{-1}$ were used to guide and confirm detector-sample position calibrations. Typically, diffraction patterns $I(Q)$ are converted to reduced structure factors $F(Q)$ and PDFs using the software PDFgetX3[37]. This conversion requires material- and measurement-dependent corrections $F(Q) = a(Q) I(Q) + b(Q)$ for unknown broadband functions $a, b$. PDFgetX3 estimates these functions ad hoc based partially on asymptotic behavior of $F(Q)$ at high and low $Q$. Due to the limited measured $Q$ range, these estimates were found to be incorrect for the XFEL data by comparison to the unpumped PDFs generated from the synchrotron reference (accounting for different detector properties etc.). Functions $a$ and $b$ were instead found by comparison to the reference unpumped measurements and fixed for all delay times. A comparison of synchrotron and XFEL PDFs is shown in Supplementary Figure 16.

To normalize the diffraction patterns by the average probe X-ray fluence, it is assumed that the total scattering intensity measured is constant with pump-probe delay. This accounts for both X-ray probe intensity and any change in sample volume due to the possibility of some sample ablation. This self-normalization procedure was verified by comparing the lowest physical PDF peak (Cu-S / Ir-S Fig. 1d) with pump-probe delay which did not meaningfully evolve (as expected, Supplementary Figure 17).

*PDF Fitting*

PDF structure fitting was carried out using the diffpy.cmi Python package wrapping the PDFFIT2 engine[38]. Structural models were optimized to minimize the root-square-difference between experimental $G_{exp}(r)$ and numerical $G_{calc}(r)$ PDFs, as quantified by residual $rw(\%) = 100 \times \sqrt{\int [G_{exp}(r) - G_{calc}(r)]^2 dr / \int [G_{exp}(r)]^2 dr}$. Local and average PDF length scales are separated at 8.8 Å as this threshold does not cut through any PDF peaks. The unpumped reference PDF was defined as the PDF measured with -20 ps pump-probe delay, which could be scaled in magnitude during the fitting to account for the fraction of unpumped material. For fits of the average structure, no parameter constraints were applied beyond those required by the crystal symmetry. Atomic displacement parameters presented represent variance in atomic positions, typically denoted as $U_{iso}$, and not the alternative $B$ value given by $8\pi^2 U_{iso}$ also presented in literature.

**Acknowledgements**

Work at Brookhaven National Laboratory was supported by the U.S. Department of Energy, Office of Science, Office of Basic Energy Sciences, under Contract No. DE-SC0012704. Use of the Linac Coherent Light Source (LCLS), SLAC National Accelerator Laboratory, is supported by the U.S. Department of Energy, Office of Science, Office of Basic Energy Sciences under Contract No. DE-AC02-76SF00515. This work was supported by the National Institutes of Health grant S10 OD023453. Work at Argonne National Laboratory (sample synthesis and characterization) was supported by the US Department of Energy Office of Science, Basic Energy Sciences, Materials Science and Engineering Division. S.D.M. and P.G.E. gratefully acknowledge support from the U.S. DOE Office of Science under grant no. DE-FG02-04ER46147 and from the US NSF through the University of Wisconsin Materials Research Science and Engineering Center (DMR-2309000 and DMR-1720415). We would like to thank Andrey A. Yakovenko and Uta Ruett for help with reference synchrotron measurements of equilibrium CuIr$_2$S$_4$. The measurements were carried out at the Advanced Photon Source (APS) beamline 11-ID-C, which was supported by the U. S. Department of Energy, Office of Science, Office of Basic Energy Sciences, under Contract No. DE-AC02-06CH11357.

# Resolving Length Scale Dependent Transient Disorder Through an Ultrafast Phase Transition

Supplementary Information

Jack Griffiths[1], Ana Flávia Suzana[1], Longlong Wu[1], Samuel D. Marks[2], Vincent Esposito[3], Sébastien Boutet[3], Paul G. Evans[2], J. F. Mitchell[4], Mark P. M. Dean[1], David A. Keen[5], Ian Robinson[1], Simon J. L. Billinge[1,6], Emil S. Bozin[1]

[1] *Condensed Matter Physics and Materials Science Division, Brookhaven National Laboratory, Upton, NY 11973, USA*
[2] *Department of Materials Science and Engineering, University of Wisconsin, Madison, WI 53706, USA*
[3] *SLAC National Accelerator Laboratory, 2575 Sand Hill Road, Menlo Park, California 94025, USA*
[4] *Materials Science Division, Argonne National Laboratory, Lemont, Illinois 60439, United States*
[5] *ISIS Neutron and Muon Source, STFC Rutherford Appleton Laboratory, Harwell Campus, Didcot, Oxfordshire OX11 0QX, United Kingdom*
[6] *Department of Applied Physics and Applied Mathematics, Columbia University, New York, New York 10027, USA*

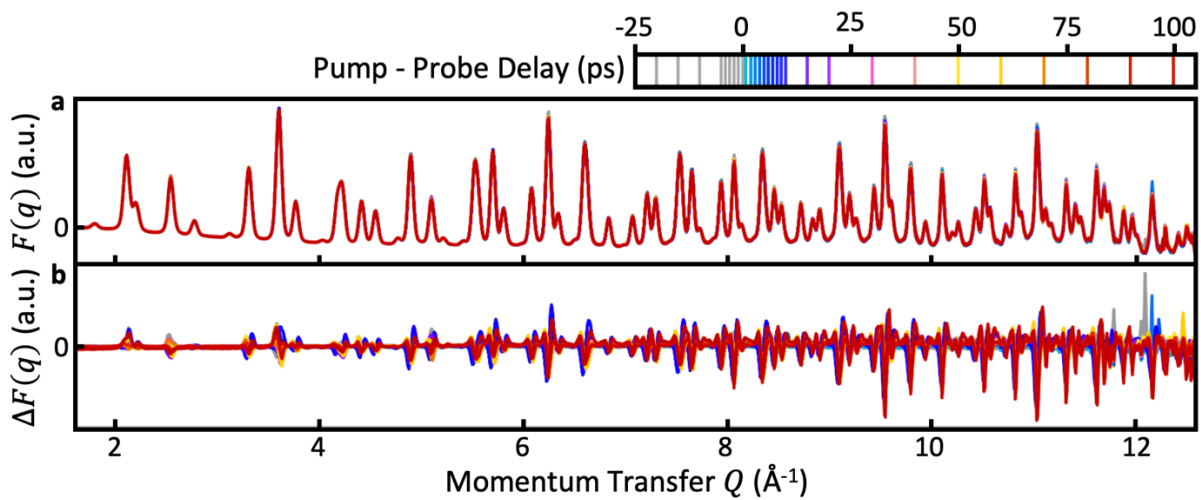

**Supplementary Figure 1 | 300 K Pump Response.** a) Reduced structure factor $F(Q)$ with varying pump-probe delay at 300 K using 27μJ of pump energy. All negative delay signals (unpumped measurements) are coloured grey. b) $\Delta F(Q)$ subtracting the average of {-20,-15,-10} ps to emphasise differences due to laser pump. Only a mild heating response (seen most clearly by the drop in high $Q$ peak intensities) is seen. Any peak shift due to lattice shifting is much smaller than the peak width.

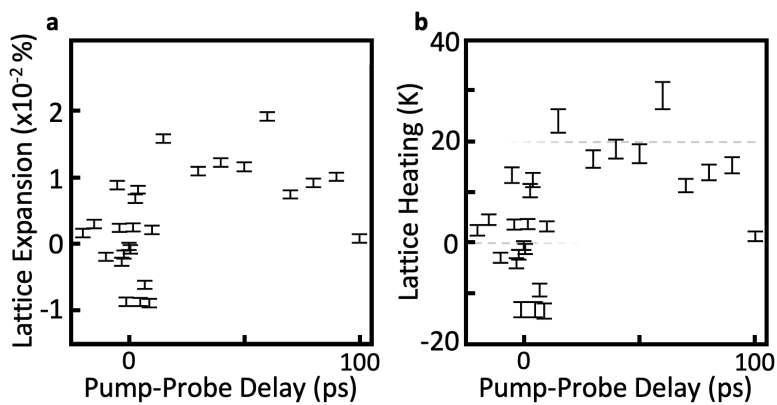

**Supplementary Figure 2 | Room Temperature Lattice Heating.** a) Thermal lattice expansion of the cubic metallic phase at 300 K under 27μJ of pump laser energy, extracted using the shift of Bragg diffraction peaks. Uncertainties are derived from maximum likelihood methods, but do not account for systematic effects while converting 2D diffraction images to 1D patterns that may increase uncertainty on these small shifts. b) Using APS synchrotron measurements of the same powder at equilibrium at 250 K and 300 K, this metallic phase is calculated to have a linear thermal expansion coefficient of $(6.7 \pm 0.6) \times 10^{-6}$ K$^{-1}$. This converts the lattice expansion into approximately 20 K of laser heating. Note that the heating onset is delayed by approximately 10 ps in accordance with the two-temperature model of lattice heating. During the 150 K pump-probe experiments, 41 μJ of pump energy was used. This would have led to approximately 30 K of lattice heating, leaving the sample still approximately 50 K below the thermal transition temperature.

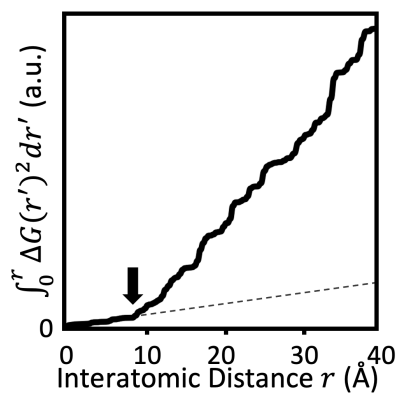

**Supplementary Figure 3 | PDF Length Scale Ranges.** Cumulative integral of 150 K $\Delta G(r)$ at 100 ps (Fig. 2d). This function is approximately piece-wise linear with an abrupt change in gradient at 9 Å (arrow) encoding a similarly abrupt increase in the magnitude of $\Delta G(r)$ past this distance. Dashed line is a guide to the eye continuing the low $r$ line. If $\Delta G(r)$ was dominated by peak shifting due to lattice heating, this cumulative integral would approximate a smooth parabola rather than two linear segments.

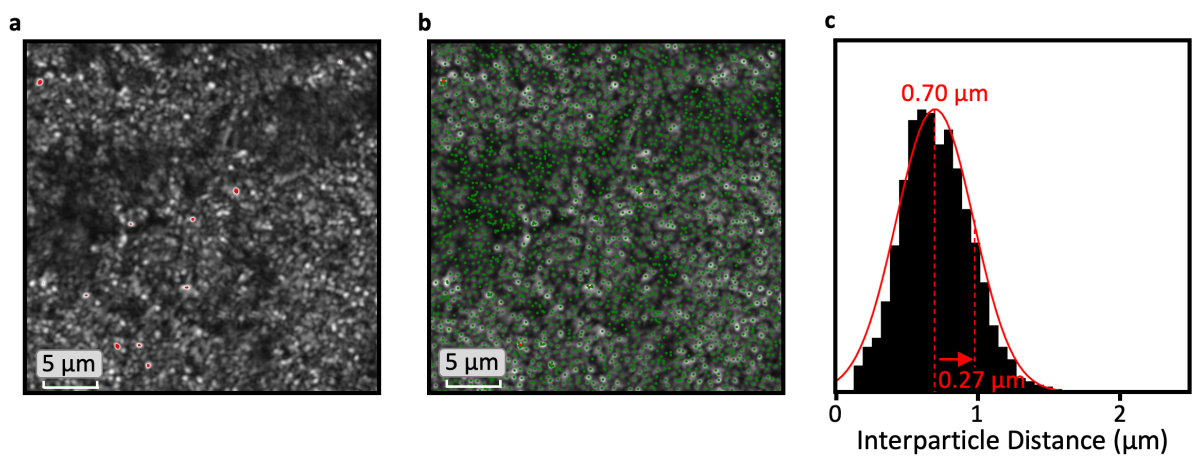

**Supplementary Figure 4 | Confocal Image of Powder CIS.** a) Confocal microscope image of the powder CIS sample spread in a thin layer on a glass slide. Red points indicate the saturation of the microscope detector. b) The center of each locally bright region, each indicating a powder particle, is marked with a green dot. c) A histogram of distances between each particle and its three nearest neighbors. This is normally distributed with a small shoulder at smaller distances due to minor artefacts in the particle detection (such as around the saturated pixels). In the limit of close packing, this gives particles with a size of (0.7 ± 0.3) μm.

## Supplementary Note 1: Calculating Difference PDFs from Dimer Breaking

To calculate the difference PDF $\Delta G(r)$ over a wider $r$-range from breaking a single Ir-dimer, a large box model of low temperature CIS was formed from a 12 x 12 x 12 supercell of the $P\bar{1}$ unit cell described in the literature[1]. This model contains 27648 Ir atoms. A single pair of dimerized Ir, at an $Ir^{4+}[1]$ and a neighboring $Ir^{4+}[4]$ position (using the literature notation provided in the cited reference) were moved apart by 0.4 Å along the $[1\bar{1}0]$ axis. Each atom was moved an equal distance from their center point. PDFs were calculated for both the unperturbed and altered structures and from these the difference PDF, $\Delta G(r) = G(r)$ [altered] – $G(r)$ [unperturbed]. This

$\Delta G(r)$ consists of a series of delta functions which are positive or negative depending on whether the altered structure has added or removed correlations at a given distance, respectively. These delta functions were then broadened by convoluting $\Delta G(r)$ with a Gaussian function with a width chosen to mimic the resolution of the experimental data. The same procedure was used to calculate $\Delta G(r)$ for a model where the $Ir^{4+}$ ions in *every* dimer were moved apart (both the $Ir^{4+}$[1] – $Ir^{4+}$[4] and $Ir^{4+}$[2] – $Ir^{4+}$[3] dimers) to completely remove the dimerization within the P$\bar{1}$ structure. The two $\Delta G(r)$ were then scaled such that the intensities of the strong $\Delta G(r)$ peaks below $r$ ~ 9 Å were approximately the same as those of the experimental $\Delta G(r)$ at short times. Scale factors of 4320 and 0.55 were used for the single dimer and all dimer breaking models, respectively. The ratio of these scale factors is approximately inversely proportional to the ratio of broken dimers in each model. The difference in correlations between the single dimer model and the unperturbed model become progressively smaller at longer distances; this converts to a calculated $\Delta G(r)$ containing significant noise at high-r because these correlations are multiplied by r when forming $\Delta G(r)$. These two calculated $\Delta G(r)$ are shown in Supplementary Figure 5 where they are compared with an averaged experimental $\Delta G(r)$ which is the PDF data taken at 4, 5 and 6 ps minus data recorded at -20, -15 and -10 ps.

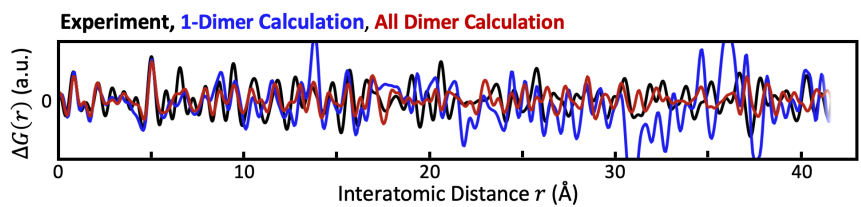

**Supplementary Figure 5 | Difference PDFs from Big Box Modelling.** Calculation of $\Delta G(r)$ from a large supercell of the refined average P$\bar{1}$ structure generated by removing a single Ir-dimer from the structure (blue line) or by removing all Ir-dimers (red line) compared with a representative short-delay time $\Delta G(r)$ (black line). Calculations are scaled by 4320x (blue) and 0.55x (red).

**Supplementary Note 2: Estimating Uncertainty on Dimer Suppression Factor**

The fraction of Ir dimers (probed by the X-ray pulse) that are suppressed by the pump laser is estimated by comparing the difference PDF dimer removal signature at 3.5 Å between the pumped transition and the equilibrium thermal transition where all dimers are removed. In figure 3, the pumped transition signature used for this purpose averages over all positive pump-probe delays. The reference PDF of dimerized CIS is defined by averaging the unpumped 150 K PDFs at -20, -15 and -10 ps pump-probe delay. Here, uncertainty on the normalization of each XFEL PDF is propagated through to the dimer suppression factor.

The first physical PDF peak (at ~ 2.3 Å, Cu-S / Ir-S) is not expected to change with pump-probe delay. The nearest neighbor distances represented by this peak reflect the strongest bonds in the system which are known to be inert even in case of substantial electron doping (e.g. by substituting Cu with Zn)[2]. However, the height of this peak does vary between PDFs by a few percent (Supplementary Figure 6a,b). These heights can be converted into a surrogate probability density function for the 'mis-scaling' factor on any given PDF by replacing each height value with a Gaussian function of finite width (Supplementary Figure 6c). A width of 0.006 is chosen as the smallest width resulting in a function that peaks once only. The function is scaled to give a unity expectation value. Repeatedly generating rescaled versions of the PDFs using samples from this distribution provides measures of uncertainty for both pumped PDFs and the reference dimerized (unpumped) PDF. In Supplementary Figure 7a and Figure 3a, each point of the difference PDF is a normal distribution (with unity height) whose width is set by this uncertainty.

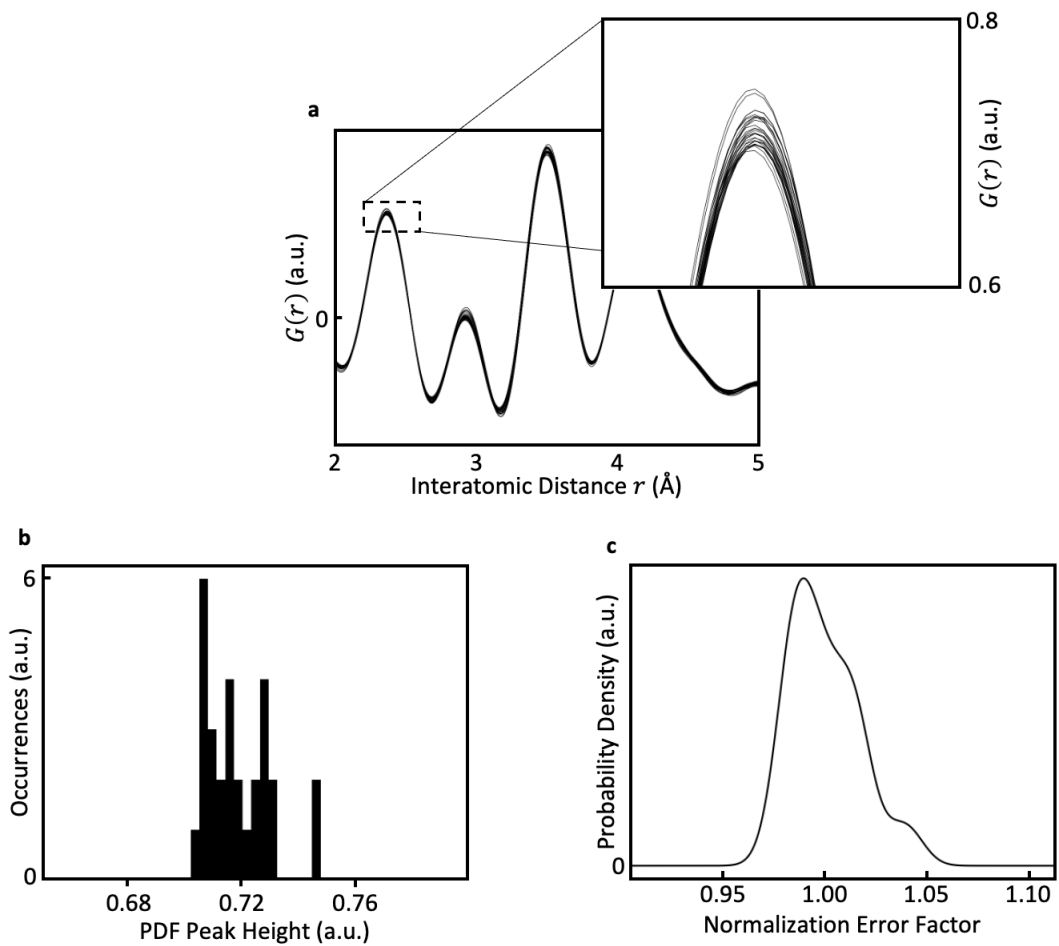

**Supplementary Figure 6 | PDF Normalization Uncertainty.** a) All PDFs overlayed from the pump-probe experiment. The lowest physical peak varies in peak height by a few percent (inset) despite being expected to be constant. b) A histogram of these PDF peak heights. c) A Probability Density Function of the PDF Normalization Error Factor generated by replacing each PDF peak height sample with a Gaussian function of width 0.006. The resulting smooth function is then scaled to a unity expectation value.

If we define the estimated dimer signature fraction to be the difference in scale between the peak of the dimer removal signature between the pumped and equilibrium transitions, we can use the generated sets of rescaled PDFs to generate a probability density function for this fraction

(Supplementary Figure 7b, Fig. 3b). This provides a value of $29.4^{+1.0}_{-0.5}\%$ for the central 68% probability interval.

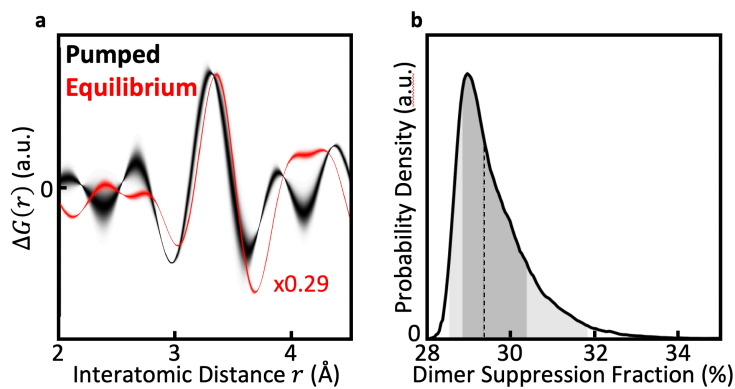

**Supplementary Figure 7 | Dimer Suppression Fraction.** a) Experimental $\Delta G(r)$ for both the pumped (black) and thermally driven equilibrium (red) transitions, showing the dimerized PDF subtracted from the un-dimerized PDF, over the low $r$ range containing the W-signature centered at 3.5 Å of strong dimer removal. The pumped data averages over all positive pump-probe delay times with line thickness indicating uncertainty. The equilibrium data is scaled by x0.29 so that the signatures approximately match in scale. b) Uncertainty on PDF normalization is propagated to a Probability Density Function representing the scaling between the pumped and equilibrium dimer suppression signatures. Dashed line indicated the median value of 29.4%. Shaded areas indicate the central 68% and 95% probability intervals.

**Supplementary Note 3: Interpolating between Dimerized and Un-dimerized Unit Cells**

We want to construct a simple model of strong dimer removal to compare to the local W signature. The unpumped material is modelled using the known $P\bar{1}$ unit cell that describes equilibrium CIS at low temperature[1]. The structure of equilibrium CIS at low temperature will be denoted here as the *dimerized structure*. The structure of equilibrium CIS at high temperature will be denoted here as the *un-dimerized structure.* To model the pumped material, we interpolate between these dimerized and un-dimerized structures. To be explicit, this means the pumped model has all atoms positioned somewhere between their positions in the dimerized and un-dimerized structures. We can then compare the difference PDF curves between the pumped and unpumped material to the experimental difference curve.

A crystal structure is defined by a set of numerical parameters that will be described here. Interpolating between two structures involves interpolating these parameter values. Importantly, the numerical description of a crystal structure is not unique. This means that meaningfully interpolation first requires the parameterization of the two structures to be chosen to be as numerically similar as possible.

A simple schematic of interpolation between two fictional 2D structures is shown in Supplementary Figure 8. The parameters defining a crystal structure include three lattice vectors defining the unit cell and fractional coordinates (between 0 and 1) for the position of each atom inside the unit cell in the basis of those lattice vectors. Although in general nine variables are required to define three vectors, lattice vectors can be freely rotated (rotating the entire structure) without affecting the calculated PDF or scattering pattern. Therefore, only six variables are needed to describe the lattice vectors: three lattice vector lengths $a$, $b$ and $c$ and the angles between the lattice vectors $\alpha$, $\beta$ and $\gamma$. For the P$\bar{1}$ description of the dimerized structure, these are {$a$: 11.976 Å, $b$: 7.000 Å, $c$: 11.956 Å, $\alpha$: 90.985°, $\beta$: 108.513°, $\gamma$: 91.033°}. The un-dimerized structure is typically described with a cubic Fd$\bar{3}$m unit cell ($a, b, c$: 9.831 Å and $\alpha, \beta, \gamma$: 90°) as this is the highest symmetry parameterization. There are many other unit cells, of apparent lower symmetry, that describe the same un-dimerized structure. We search for the unit cell whose parameters are as close as possible to those describing the dimerized structure. This results in a monoclinic unit cell with lattice parameters {$a$: 12.040 Å, $b$: 6.951 Å, $c$: 12.040 Å, $\alpha$: 90.000°, $\beta$: 109.471°, $\gamma$: 90.000°}. For this set of lattice parameters, there are many distinct sets of atomic fractional coordinates that still describe the same un-dimerized structure. Again, we pick the set of atomic fractional coordinates that are as close as possible to those in the P$\bar{1}$ description of the dimerized structure.

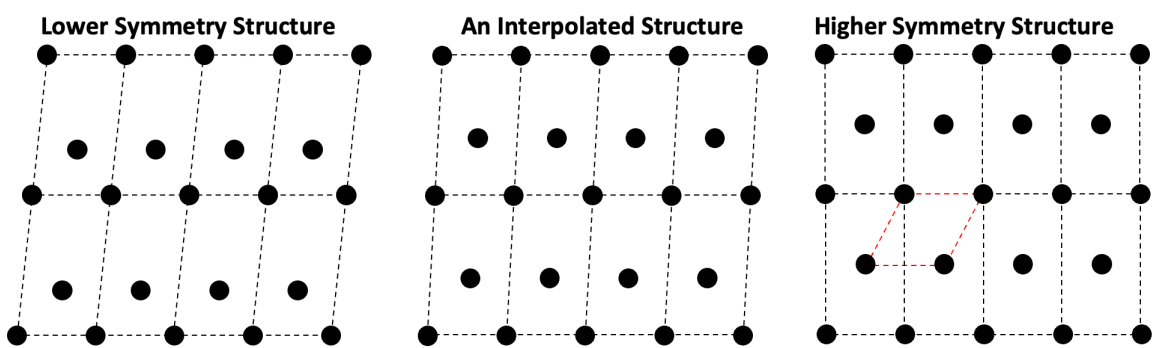

**Supplementary Figure 8 | Structural Interpolation.** Schematic example of interpolation between two fictional 2D structures. The 'Lower Symmetry Structure' has a unit cell containing 2 atoms (black dashed lines). Although the 'Higher Symmetry Structure' has a primitive unit cell containing 1 atom (red dashed lines) it can be described with a very similar unit cell to the Lower Symmetry structure (black dashed lines, containing two atoms) although with orthogonal lattice vectors and the atom in the interior of the unit cell shifted to its centre. The 'Interpolated Structure' mixes these two unit cells, making the lattice vectors of the Lower Symmetry structure more orthogonal and moving the interior atom towards the unit cell centre.

From all possible degenerate parameterizations of the un-dimerized structure, we have selected one such that the numerical values parameterizing it are as similar as possible to those parameterizing the dimerized structure. A new structure can be defined by interpolating these values. For fractional coordinates, this interpolation is trivial. However, trivially interpolating each of the six unit cell parameters can result in unintended modifications to the unit cell volume. Instead, these six parameters must be converted into full lattice vectors

$$\text{Dimerized:} \begin{bmatrix} 11.976 & 0 & 0 \\ -0.1262 & 6.9989 & 0 \\ -3.7964 & -0.2741 & 11.3344 \end{bmatrix} \text{Å}, \quad \text{Un-dimerized:} \begin{bmatrix} 12.04 & 0 & 0 \\ 0 & 6.9513 & 0 \\ -4.0133 & 0 & 11.3514 \end{bmatrix} \text{Å}$$

where each row gives a different lattice vector. Arbitrarily, in both cases the first lattice vector is placed along the x-axis and the second vector is place within the x-y plane. Interpolating these 9 values to generate a new intermediate unit cell can still suffer from this volume problem (Supplementary Figure 9). By rotating one set of lattice vectors, this problem can be either worsened or improved. We subtly rotate the un-dimerized vectors to remove the volume issue entirely:

$$\text{Dimerized:} \begin{bmatrix} 11.976 & 0 & 0 \\ -0.1262 & 6.9989 & 0 \\ -3.7964 & -0.2741 & 11.3344 \end{bmatrix} \text{Å}, \quad \text{Un-dimerized:} \begin{bmatrix} 12.04 & 0 & -0.012 \\ 0 & 6.9513 & -0.0145 \\ -4.002 & 0.0237 & 11.3554 \end{bmatrix} \text{Å}$$

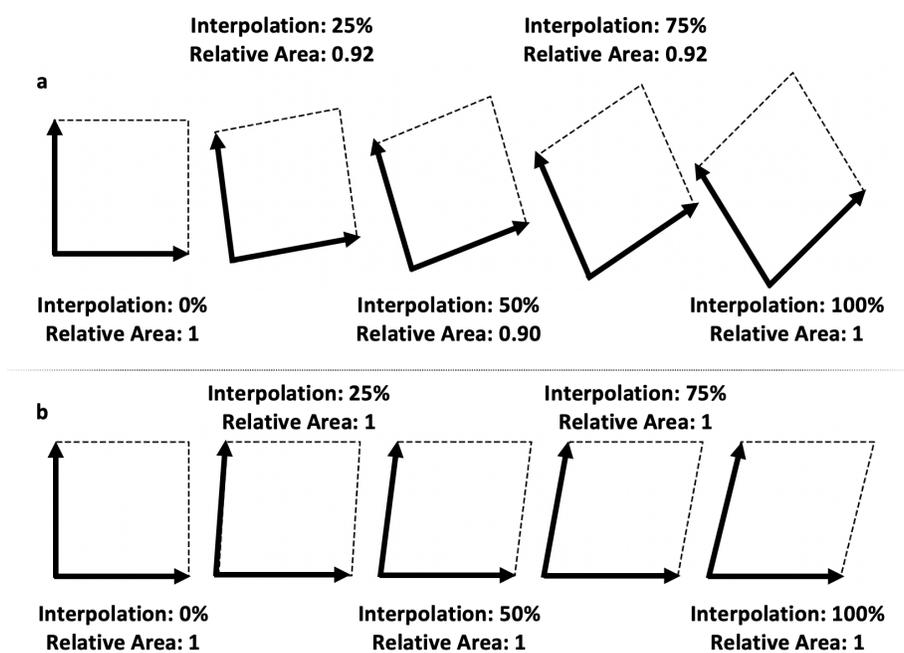

**Supplementary Figure 9 | Interpolating Lattice Vectors.** Two 2D unit cells of equal area are defined by lattice vectors that with a 90º and <90º angles respectively. A) Linearly interpolating the lattice vectors of these two cells can lead to intermediatory unit cells where area is not preserved. B) By rotating the lattice vectors of the two unit cells correctly relative to one another, area is preserved for all possible interpolation values.

We define an interpolated structure intermediate to the dimerized and un-dimerized structures with four interpolation parameters. One controls the interpolation of the lattice vectors while the others describe the interpolation of the fractional coordinates associated with each atomic species (Cu, Ir and S).

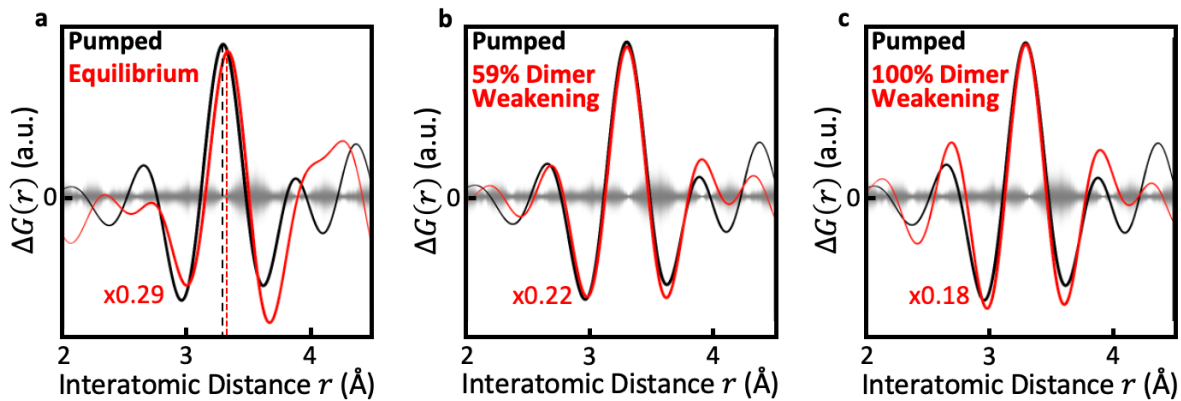

**Supplementary Figure 10 | Dimer Removal Signature Modelling.** a) Experimental $\Delta G(r)$ for both the pumped (black) and thermally driven equilibrium (red) transitions, showing the dimerized PDF subtracted from the un-dimerized PDF, over the low $r$ range containing the W-signature centered at 3.5 Å of strong dimer removal. The equilibrium data is scaled down 3.5x so that the signatures approximately match in scale. The pumped signature peaks at a smaller value of $r$ (dashed lines). Background grayscale gradient indicates the expected noise level, extracted from differences between nominally equivalent negative pump-probe delay measurements and scaled down to account for averaging of the pumped difference. b) Experimental pumped (black) and modelled (red) $\Delta G(r)$. Unpumped modelled PDF given by the known literature structure. Pumped PDF given by interpolating Ir atoms positions towards their un-dimerized configuration to drop the standard deviation of Ir-Ir distances 59%. Ir Atomic Displacement Factor (ADP) inflated by 4.5x for the pumped PDF. Modelled $\Delta G(r)$ scaled down 4.5x to match the experimental data in scale. c) Experimental pumped (black) and modelled (red) $\Delta G(r)$ where Ir atoms in the pumped structure are entirely returned to their un-dimerized configuration and Ir ADP inflated by 6.5x. Modelled $\Delta G(r)$ scaled down 5.5x to match the experimental data in scale.

To best match the average $\Delta G(r)$ for all positive pump-probe delays, the Cu, Ir and S fractional coordinates are interpolated from the dimerized towards the un-dimerized values by 0%, 61.7% and 0.5% respectively. The lattice parameters interpolate 0%. This gives a 59% drop in the spread (standard deviation) of Ir-Ir distances (Supplementary Figure 10b). The Ir ADP is inflated 4.5x between the unpumped and pumped PDFs to 0.008 Å², shifting the W signature to slightly lower $r$ due to termination effects. Within the noise level, this cannot be reliably distinguished from a model with the Ir fractional coordinates and lattice parameters fixed at 100% interpolation to remove all spread in Ir-Ir distances (Supplementary Figure 10c). The Cu and S interpolations optimize to 0 and 7% respectively in this case and the Ir ADP is inflated 6.5x to 0.013 Å².

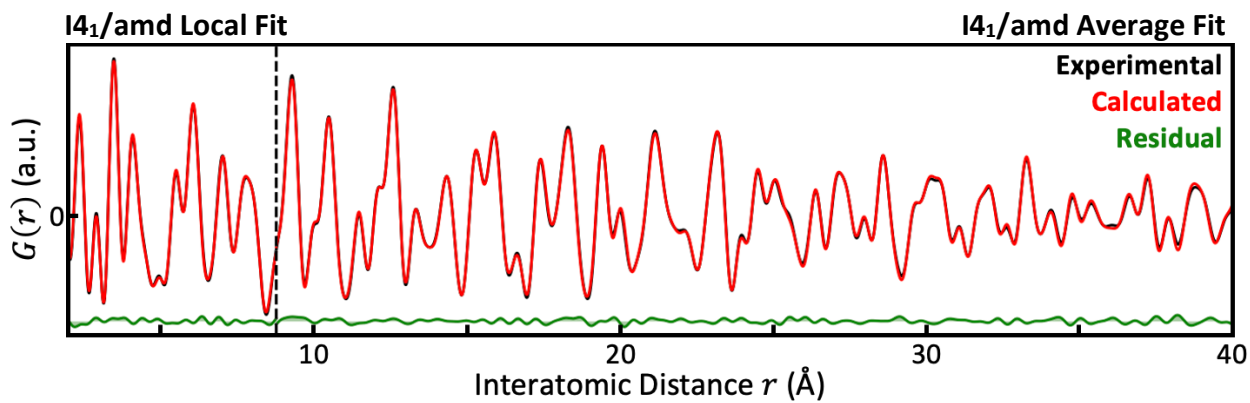

**Supplementary Figure 11 | Example Pumped PDF Modelled.** Experimental (black) and calculated (red) PDFs for 150 K pumped CIS at 100 ps pump-probe delay using the I4$_1$/amd crystal group. Independent models are applied over the local and average ranges, separated by the vertical dashed line.

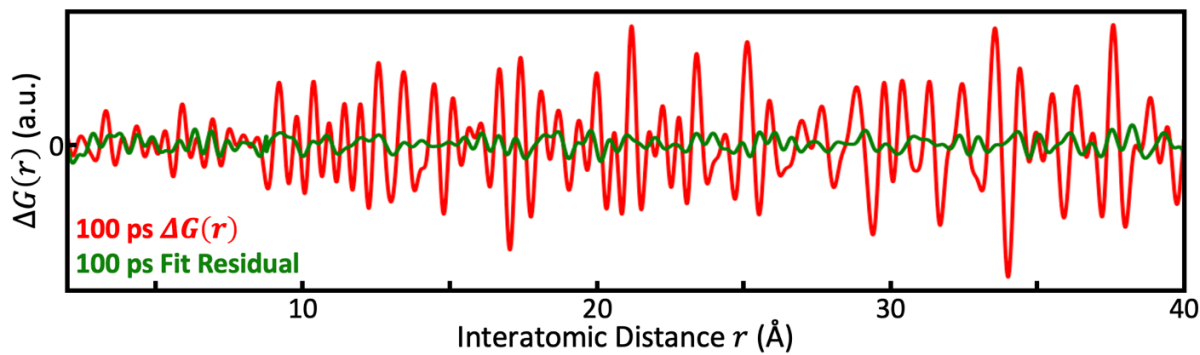

**Supplementary Figure 12 | 100ps $\Delta G(r)$ v. Fit residual.** $\Delta G(r)$ for the 150 K 100 ps pump probe delay measurement (red) as shown in Figure 2d. For comparison, the fit residual of this PDF is also shown (green) when using the I4$_1$/amd crystal group, as shown in Supplementary Figure 11. The fit residual is significantly smaller in magnitude than the effect of the pump laser, meaning that the model captures well the features of the pumped state.

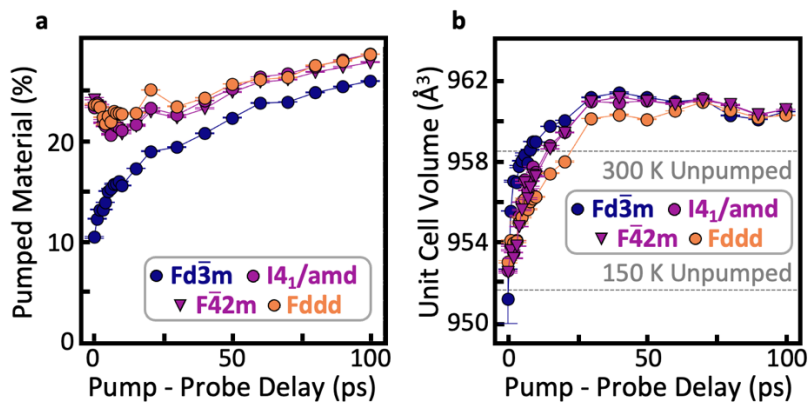

**Supplementary Figure 13 | Non-Local Structural Models.** Additional model parameters from the small box modelling of the 8.8 – 40 Å. a) The modelled pumped phase fraction with pump-probe delay. This is only

significantly different for the Fd$\bar{3}$m unit cell which struggles with underfitting at low delay. b) The unit cell volume with pump-probe delay is consistent for all models. Dashed lines indicate volumes for the 150 K and 300 K unpumped CIS material.

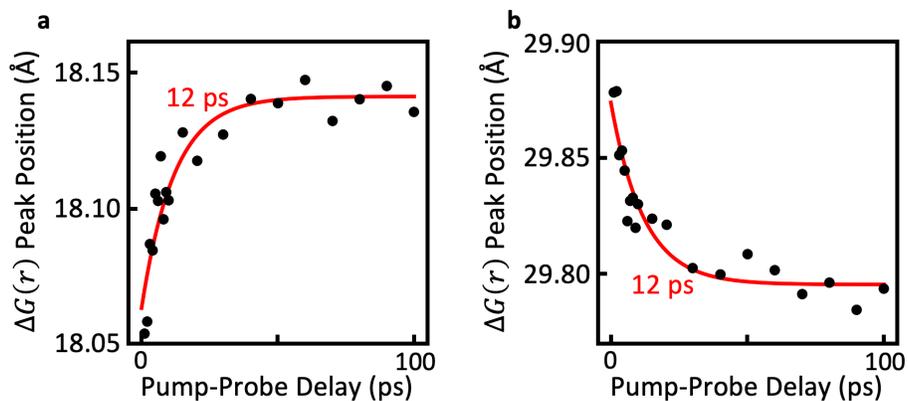

**Supplementary Figure 14 | $\Delta G(r)$ Peak Shifting.** Some peaks in $\Delta G(r)$ (Fig. 2d) shift subtly with pump-probe delay. This does not indicate simple lattice expansion, as not all peaks shift and those that do can shift to a) higher or b) lower interatomic distances. This shifting matches well to the 12 ps timescale extracted from structural modelling (red lines).

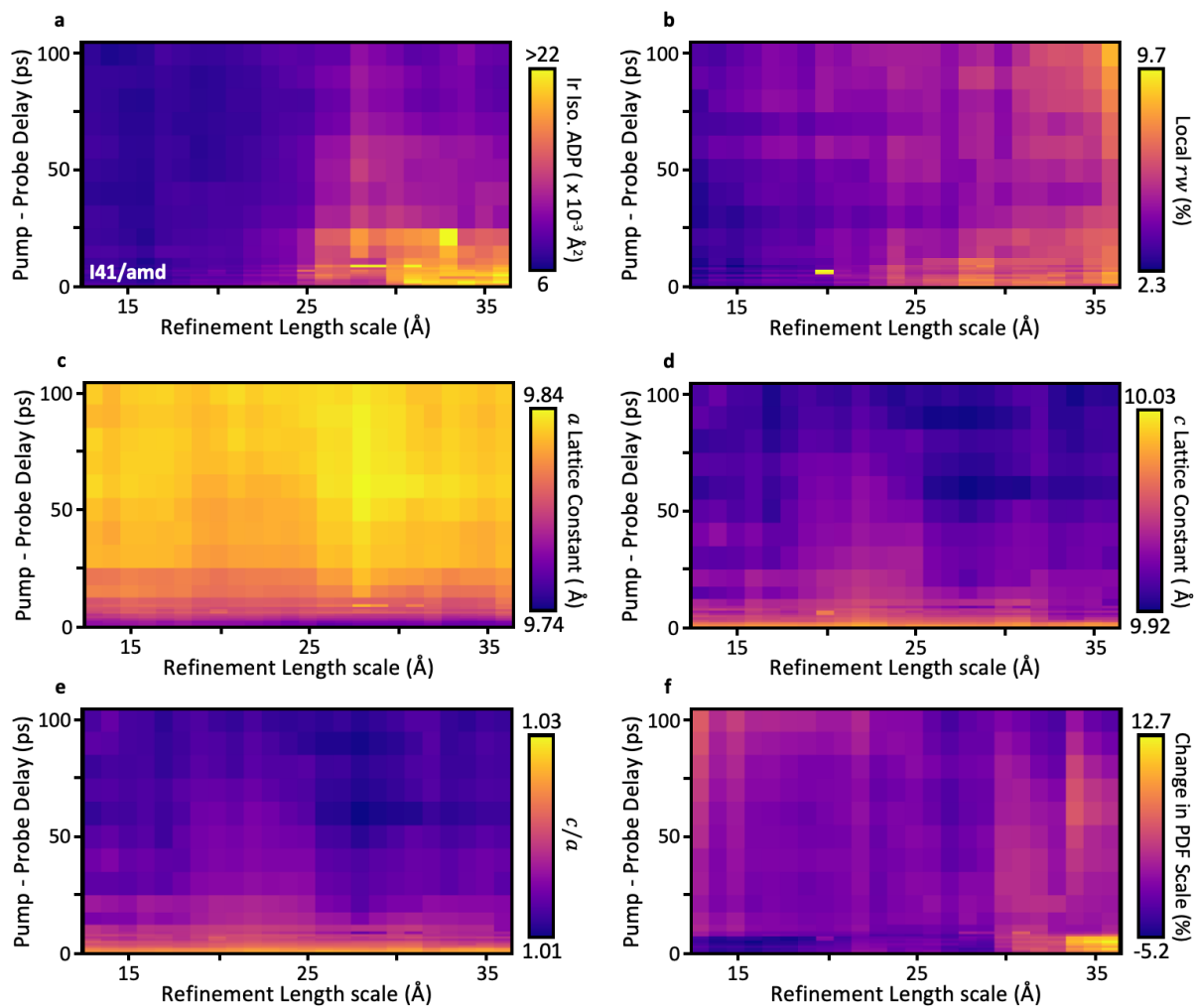

**Supplementary Figure 15 | PDF Sliding Window Refinements.** I4$_1$/amd model refined over a sliding widow of width 8.2 Å. Refinement Length Scale indicates the center of the refinement window. a) The isotropic Ir ADP as shown in Figure 4e. b) Local residual for each windowed refinement. To ensure this is not an artifact of correlated refinement parameters, the c) a lattice constant, d) c lattice constant and e) a/c ratio do not display changes correlated with the Ir ADPs. f) The scale factor used to match the experimental and modelled PDFs, normalized by the mean average for each pump-probe delay, also does not show the same length scale dependance.

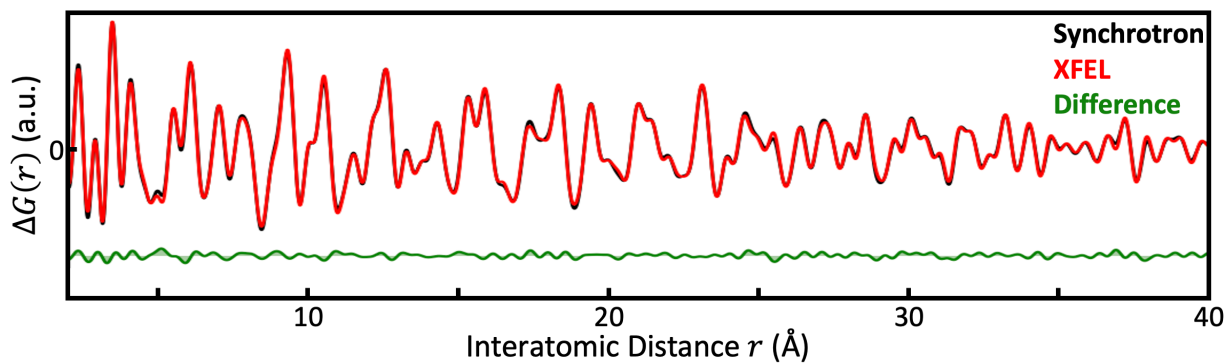

**Supplementary Figure 16 | Synchrotron and XFEL PDFs.** PDFs of unpumped CIS at 150 K taken at the APS synchrotron (black) and at the LCLS XFEL (red, data is -20 ps pump-probe delay). Green indicates the difference between the two data sets. The synchrotron data is trimmed in reciprocal space to match the momentum transfer range of the XFEL measurement. The XFEL data is corrected with $r$-dependant scaling term $\propto \exp(-0.5 \times (Ar)^2$, where $A$ is a tunable parameter, to account for the different reciprocal space resolutions of the two measurements.

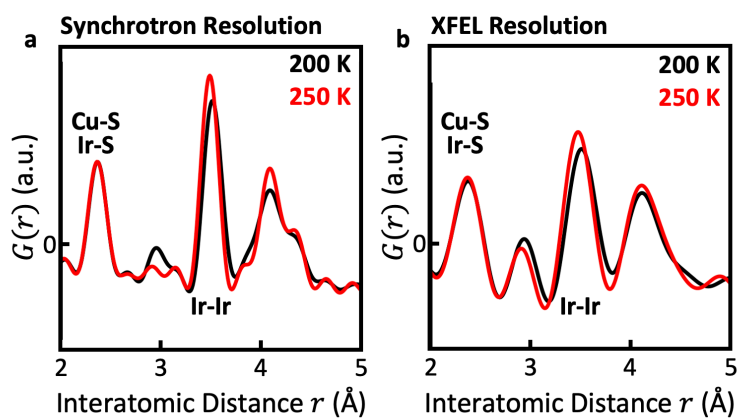

**Supplementary Figure 17 | Equilibrium Synchrotron Measurements.** a) Unpumped CIS PDF measured at the APS synchrotron above and below the transition temperature using a high maximum $Q$ of 23 Å$^{-1}$. The dimerization signature around the Ir-Ir peak is clear while the first peak (Cu-S and Ir-S) is left insignificantly changed. b) These observations hold when artificially reducing the $Q$ range of this same data to match the XFEL experiment in this work, lowering the PDF resolution.